\documentstyle[12pt,epsf,epsfig,wrapfig]{article}
\newcommand{\rar}{\rightarrow}
\newcommand{\pdup}{p_\uparrow}
\newcommand{\ddup}{d_\uparrow}
\newcommand{\ndup}{N_\uparrow}
\newcommand{\ppdup}{p + \pdup \rar \pi^0 + X}
 
\newcommand{\pimp}{\pi^- + \pdup \rar \pi^0 + X}
\newcommand{\pimd}{\pi^- + \ddup \rar \pi^0 + X}
\newcommand{\pimn}{\pi^- + \ndup \rar \pi^0 + X}

\def\Journal#1#2#3#4{#1 {\bf #2}, #3 (#4)}
\newcommand{\PLB}{Phys.\ Lett.}
\newcommand{\PAN}{Phys.\ Atom.\ Nucl.}
\newcommand{\YAF}{Yad.\ Phys.}

\begin{document}

\begin{center}
{\bfseries THE COMPARATIVE STUDY OF THE INCLUSIVE $\pi^0$ ANALYZING POWER
IN REACTIONS $\ppdup$ AND $\pimp$ AT 50 AND 40 GEV/C RESPECTIVELY}

\vskip 5mm

V.V.~Mochalov$^{1 \dag}$, \underline{S.B.~Nurushev}$^{1}$, A.N.~Vasiliev$^1$,
N.A.~Bazhanov$^2$, N.S.~Borisov$^2$, Y.M.~Goncharenko$^1$, A.M.~Davidenko$^1$,
A.A.~Derevschikov$^1$, V.G.~Kolomiets$^2$, V.A.~Kormilitsin$^1$,
V.I.~Kravtsov$^1$, A.B.~Lazarev$^2$, Yu.A.~Matulenko$^1$, Yu.M.~Melnick$^1$, 
A.P.~Meschanin$^1$, N.G.~Minaev$^1$, D.A.~Morozov$^1$, A.B.~Neganov$^2$, 
L.V.~Nogach$^1$, Yu.A.~Plis$^2$, A.F.~Prudkoglyad$^1$, A.V.~Ryazantsev$^1$, 
P.A.~Semenov$^1$, O.N.~Shchevelev$^2$, L.F.~Soloviev$^1$, Yu.A.~Usov$^2$, 
A.E.~Yakutin$^1$ \\

\vskip 5mm
{\small (1) 
{\it Institute for High Energy Physics, Protvino, Russia}\\
(2) {\it Joint Institute for Nuclear Research, Dubna, Russia}\\
$\dag$ {\it E-mail: mochalov@ihep.ru}}

\end{center}

\vskip 5mm

\begin{abstract}
Single-spin asymmetries $A_N$ in reactions $\ppdup$ and $\pimp$ 
at 50 and 40~GeV/c respectively behave in drastically different 
ways in function of transverse momentum in the central region. 
At the same time $A_N$ in the polarized proton fragmentation region 
of these reactions are practically coinciding. 
Our new data on the analyzing power at 50~GeV/c in the polarized 
proton fragmentation region in reaction $\ppdup$ confirm this conclusion 
with better statistics and coincide with our previous data at 
70~GeV/c for the same reaction.
\end{abstract}

\vskip 8mm

Our previous measurements (see Fig.~1) of the single spin 
asymmetries in reactions $\pimp (1)$ at 40~GeV/c~\cite{protvplb} 
and $\ppdup (2)$ at 70~GeV/c~\cite{protv70cent} showed that they 
behave in drastically different ways in function of transverse momentum 
in the central region. 

\begin{figure}[b!]
\begin{center}
\begin{tabular}{cc}
\mbox{\epsfig{figure=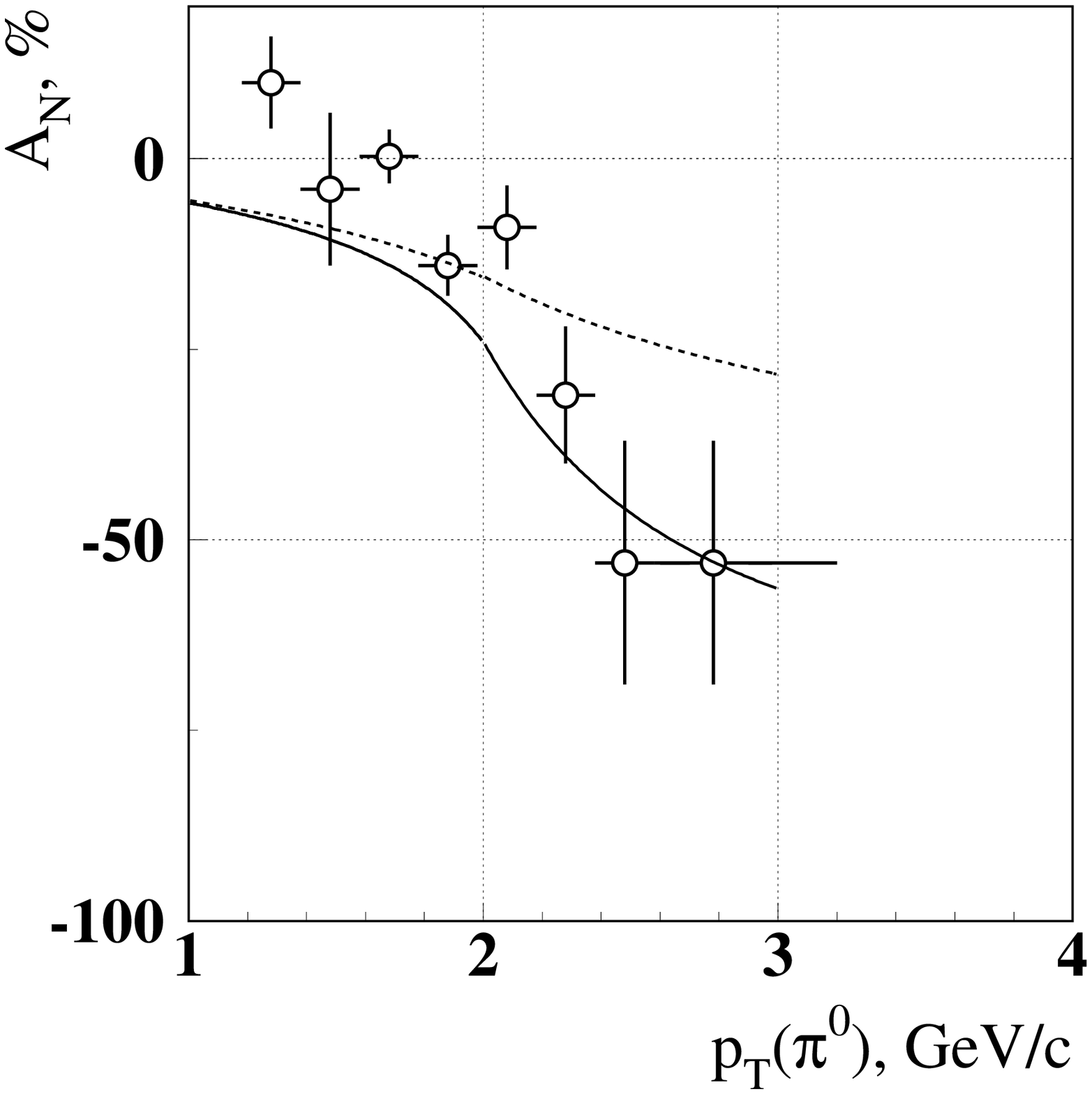,width=0.28\textwidth}}& 
\mbox{\epsfig{figure=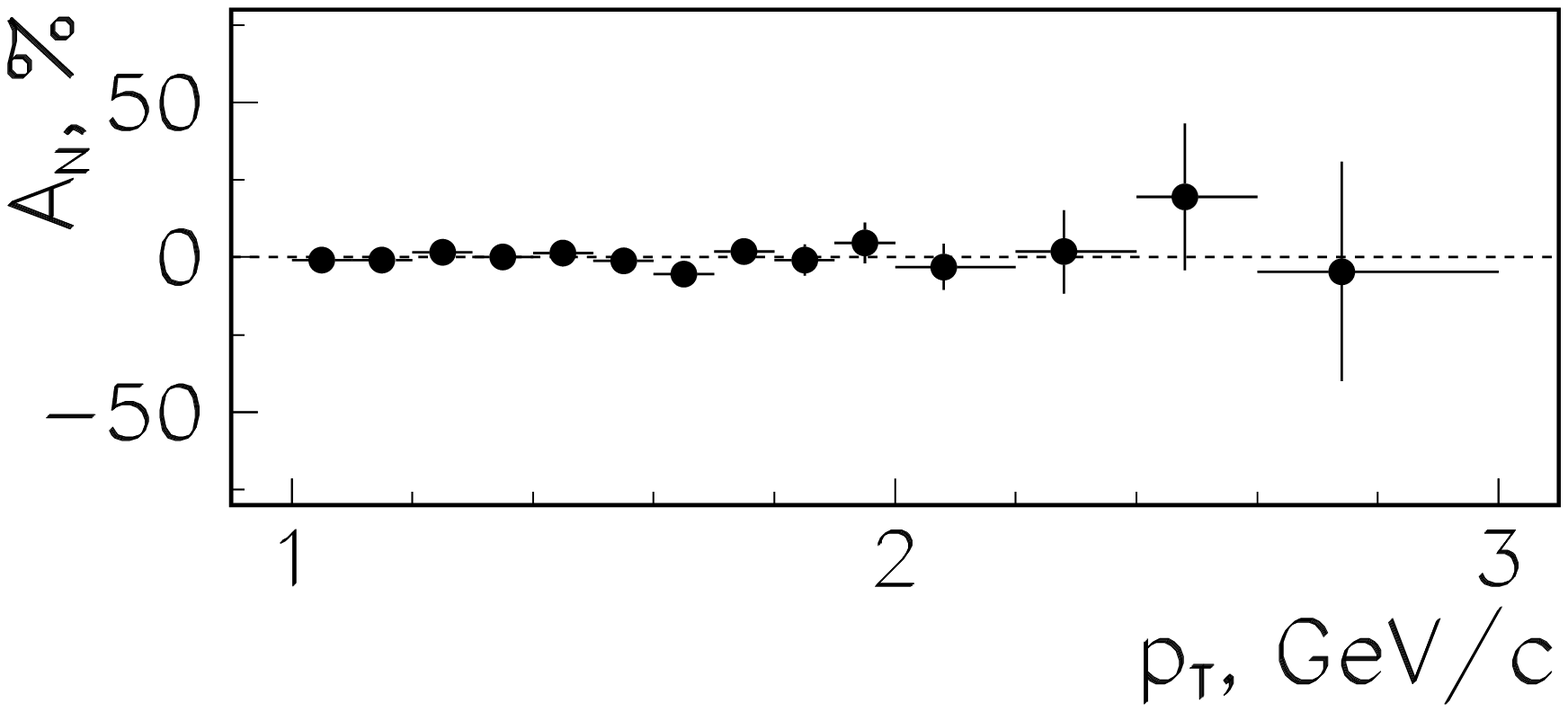,width=0.65\textwidth}}\\ 
{\bf(a)} & {\bf(b)}\\
\end{tabular}
\end{center}
\vskip -5mm
{\small{\bf Figure 1.} 
Analyzing power of the reactions $\pimn$ {\bf (a)} and $\ppdup$ {\bf (b)} at the central region at 40 and 70~GeV/c respectively.}
\end{figure}

The analyzing power of reaction (1) is close to zero around the momentum 
transfer $p_T$=1~GeV/c and then increases with growth of $p_T$ up to 
40\% for $p_T>$2.2~GeV/c. The same behavior was found in 
reaction $\pimd$ at the same kinematical region~\cite{protv_yaf}. 
At the same time $A_N$ for reaction (2) at 70~GeV/c is compatible with zero 
in the central region for the same domain of the transverse momentum. 
We may think about the following sources of the discrepancy. 
First one is related to the difference in the initial momentum of the 
incident particles, namely, 40 and~70 GeV/c. This argument does 
not work for the following reason. As it's well known, one half of the 
incident hadron energy is carried by quarks and another one half is carried 
by gluons. Therefore the momentum of the interacting incident quark is 
10~GeV/c in the case of pion beam and around 12~GeV/c in the case of proton 
beam. Assuming that the incident quark interacts with the constituent quark 
of the mass around 0.3~GeV we estimate the initial energy of the quark 
interaction in the center of mass system as 2.4~GeV for reaction (1) and 
2.7~GeV for reaction (2). Such a small difference in the interaction energy 
should exclude the big difference in the spin effects in the reactions 
under discussion. As we show later our new result on $A_N$ for reaction 
(2) at 50~GeV/c experimentally confirms such conclusion for the beam 
fragmentation region. The second possibility for difference in analyzing 
power  might be the existence of the antiquark in pion and the possible 
role of the annihilation process. We are not aware of any theoretical 
judgments about this subject.

\begin{figure}[b!]
\begin{center}
\begin{tabular}{cc}
\mbox{\epsfig{figure=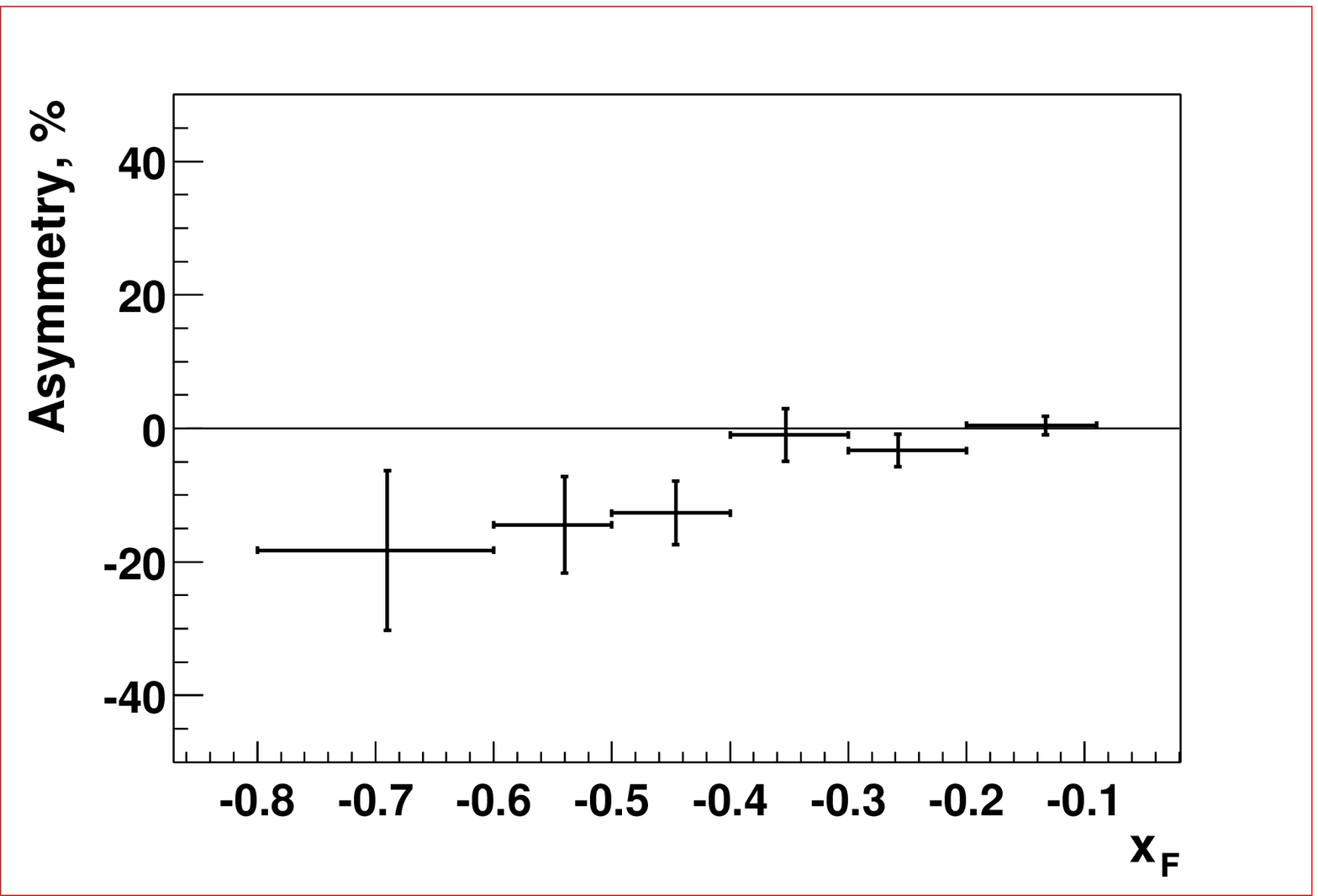,width=0.4\textwidth}}& 
\mbox{\epsfig{figure=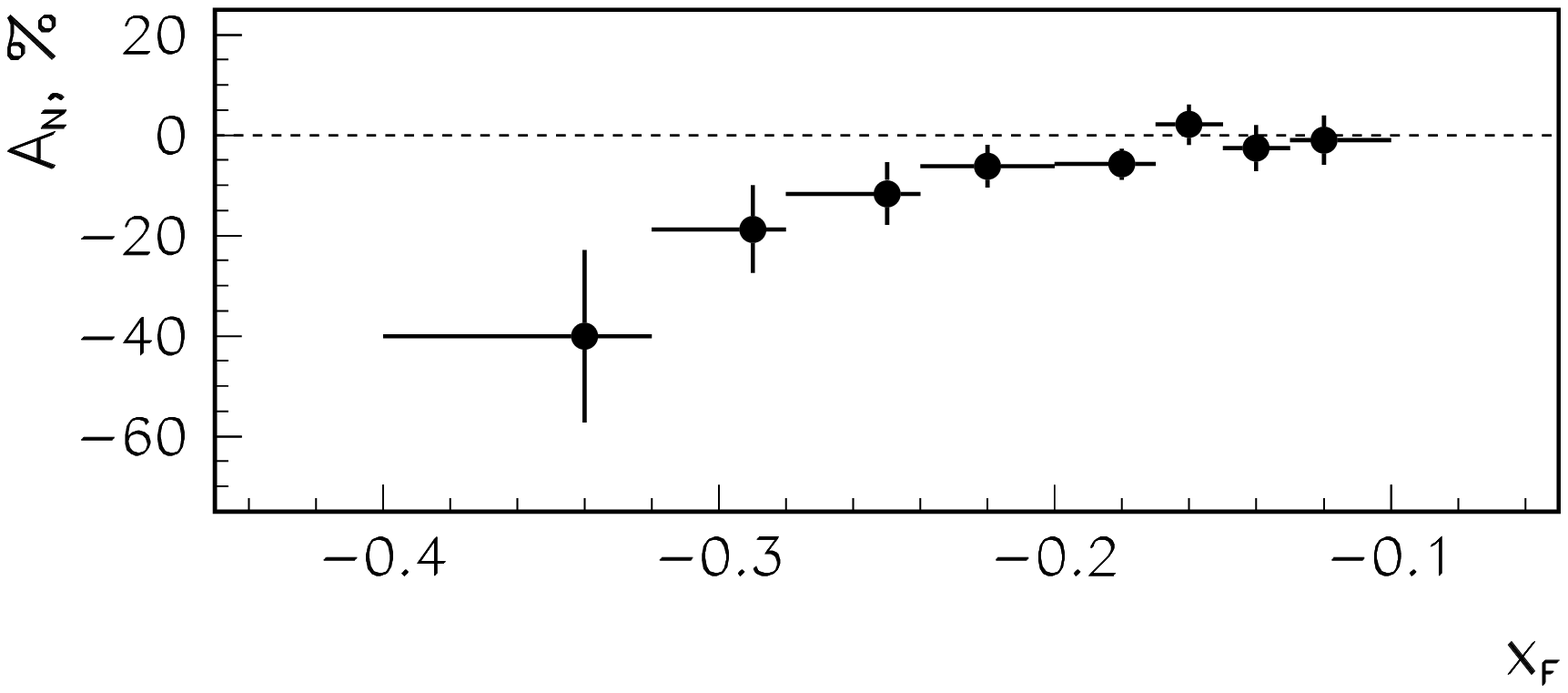,width=0.55\textwidth}}\\ 
{\bf(a)} & {\bf(b)}\\
\end{tabular}
\end{center}
\vskip -5mm
{\small{\bf Figure 2.} 
Analyzing power of the reactions $\pimp$ {\bf (a)} and $\ppdup$ {\bf (b)} at 
the polarized target fragmentation region at 40 and 70~GeV/c respectively.}
\end{figure}

The next discovery of the PROZA Collaboration, presented in Fig.2,
is relevant to the single spin asymmetries in reactions (1) and (2) in the 
polarized proton fragmentation region. 
The asymmetry in the reaction (1) is close to zero in the interval 
$0<-x_F<0.4$, then increases with growth of the $|x_F|$ reaching the value 
around 30\% at $|x_F|$=0.7~\cite{protv40}. Similar behavior is 
illustrated by the reaction (2)~\cite{protv70back}. So we do not see 
the flavor dependence of the asymmetry in contrast to the data for 
those reactions at the central region.

The goal of this article is to present our new data for reaction (2) with 
better statistics, but at the initial proton momentum 50~GeV/c, 
which corresponds to the quark energy in c.m.s around 2.2~GeV.

\begin{figure}[t!]
\epsfig{figure=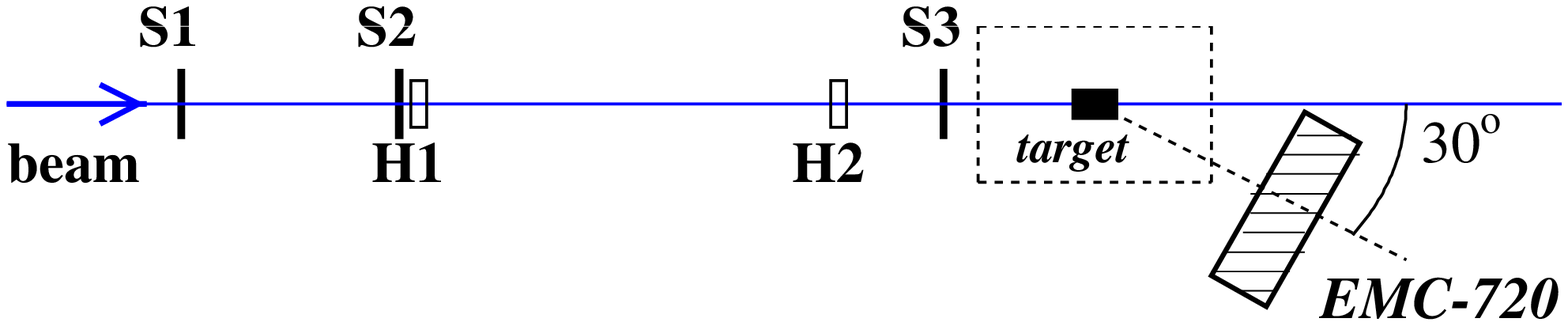,width=0.8\textwidth} 
\vskip -5mm
{\small{\bf Figure 3.} 
Experimental Setup PROZA-M. S1-S3 -- trigger scintillation 
counters; H1-H2 -- hodoscopes; $EMC-720$ -- electromagnetic 
calorimeter; $target$ -- polarized target.
}
\end{figure}

The layout of the experiment PROZA-M is presented in Fig.3. 
The proton beam of momentum 50~GeV/c extracted by curved 
mono-crystal~\cite{beam} comes from the left side, passes through the 
scintillation counters S1-S3, hodoscopes H1, H2 and strikes the polarized 
proton propane-diol target (PPT).
Specific features of the PPT are the fairly high target polarization (90\%), 
the long polarization life time and sufficiently large target length which 
was used in the frozen spin mode.~\cite{target}. 
The photons emitted from target are detected by the electromagnetic 
calorimeter EMC-720, consisting of 720 lead glass counters 
packed as 30$\times $24 matrix. Cell sizes are 
38.1$\times$38.1$\times$450 mm$^3$ (18~$X_0$). 
It is installed under angle 30$^\circ$ to the beam direction at the 
distance $l=$2.16~m from the center of the PPT. The dashed box around the 
PPT denotes the unique magnet carrying two functions: building up the 
target polarization and holding it during the data taking. 
The PROZA setup is described in detail somewhere~\cite{setup}.

EMC was calibrated by wide electron beam of 5~GeV/c using inverse matrix 
method. Sensitivity of the ADC channels is about 2.2~MeV/channel. 
Additional calibration using $\pi^0$-mass during data taking was used to 
monitor EMC energy stability in time with accuracy 0.1\%. 

Trigger requires the coincidence of signals from three scintillation counters, 
at least one hit in each plane of hodoscopes and total deposited energy 
in EMC $\Sigma E>2$~GeV. The DAQ system includes the registers for 
hodoscopes, 12 bits ADC for EMC, scalers, the read-out processor on the 
base of processor MC68030. In average 700 events per spill were registered. 
During 10 days data taking $5\cdot 10^7$ events were accumulated.

\begin{figure}[b!]
\begin{center}
\begin{tabular}{cc}
\mbox{\epsfig{figure=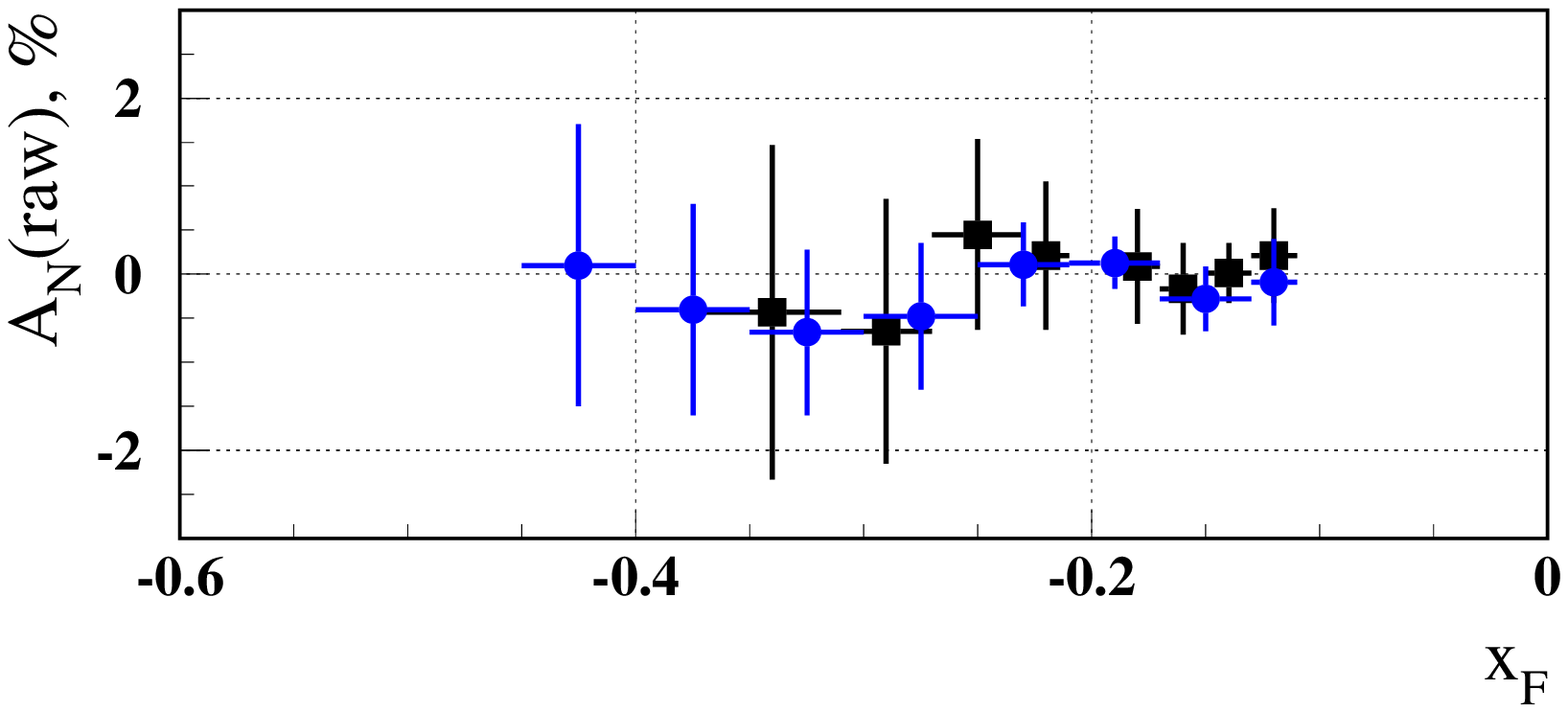,width=0.49\textwidth}}& 
\mbox{\epsfig{figure=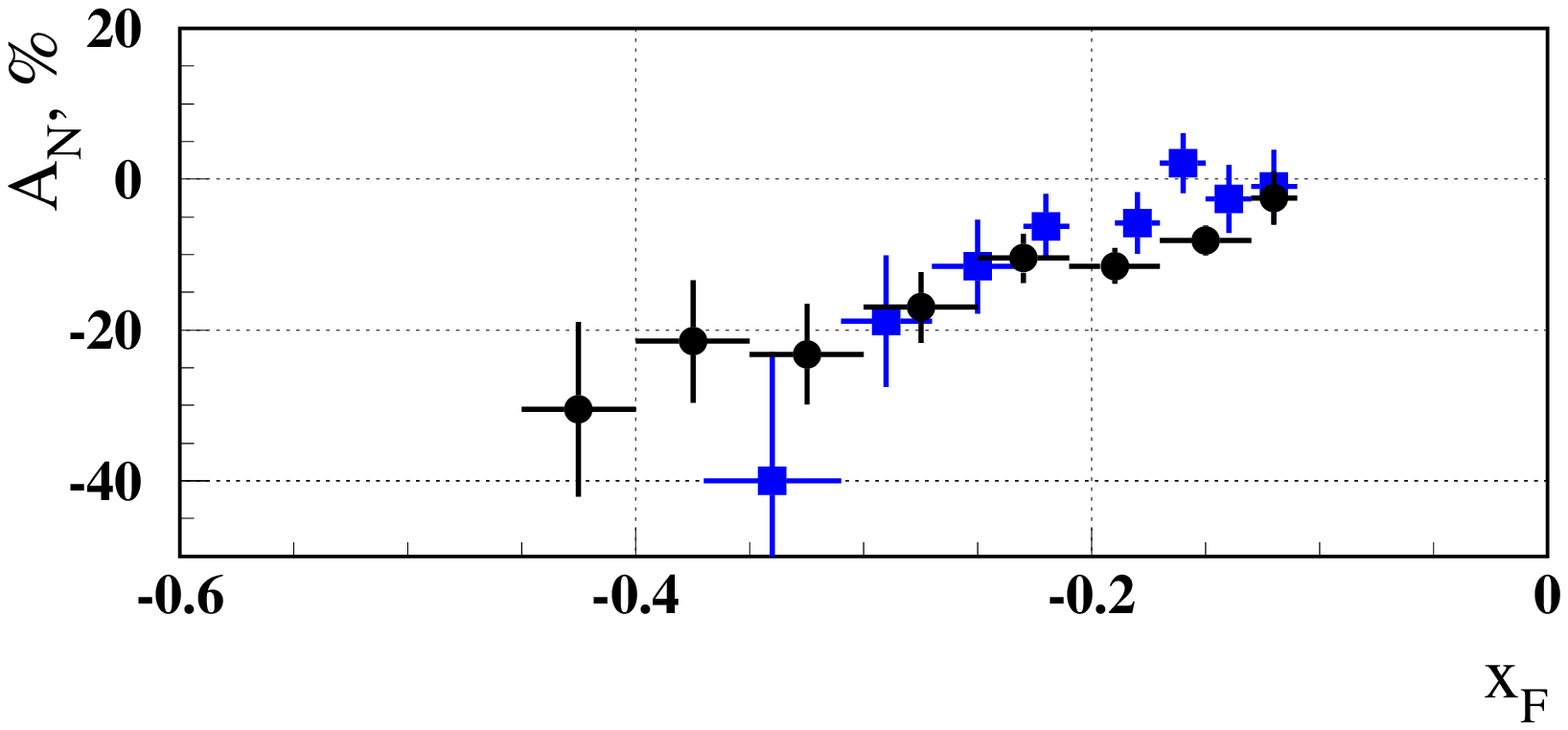,width=0.49\textwidth}}\\ 
{\bf(a)} & {\bf(b)}\\
\end{tabular}
\end{center}
\vskip -5mm
{\small
{\bf Figure 4a.} False raw asymmetry for different sets of data \\
{\bf Figure 4b.} $A_N$ in the reactions $\ppdup$ at 
the polarized target fragmentation region at 50~GeV/c (circles) 
and 70~GeV/c (squares).}
\end{figure}

For shower reconstruction it is required that at least 5 cells among 9 
central (3$\times 3$) were activated; energy deposit in the central counter 
should be at least 100~MeV. 
For reconstruction of the $\pi^0$ the photons in the energy region 0.5-5~GeV 
were used. Additional procedures were implemented to reconstruct actual 
photon energy and coordinate:
\begin{enumerate}
\item
\vspace*{-0.4cm} 
The dependence of the reconstructed photon energy on the real initial 
photon energy\cite{protv70cent}. This correction was of order of 10\%.
\item
\vspace*{-0.4cm} 
The dependence of the reconstructed photon energy and coordinate on its 
inclination angle\cite{sol1,sol2}. The energy correction was of 
order 5\%. The coordinate correction is 2-3 cm for 15$^\circ$ 
gamma inclination angle.
\end{enumerate}

After corrections the reconstructed $\pi^0$ mass was consistent with its 
table mass within precision less than 1\% in whole kinematical range. 

The raw asymmetry was calculated by usual way normalizing the counting 
rate to the events outside of the $\pi^0$-mass region. In order to check 
that the false asymmetry is zero for the fixed target polarization 
the vents were divided in two groups with almost equivalent statistics. 
Using these two groups we calculated the false asymmetry. 
Such procedure was applied to both sign of the polarization target 
independently. The results for such false asymmetries are presented in Fig.4a. 


New results for the reaction $\ppdup$ in the polarized target fragmentation 
region at 50~GeV are shown in Fig.4b. $A_N$ in the inclusive $\pi^0$ 
production at polarized target fragmentation region increases by magnitude 
with growth of $|x_F|$ and achieves $-(20.4 \pm 3.3)\%$ at $-0.45<x_F<-0.25$.
These data are consistent with our previous measurement of the analyzing 
power for the same reaction and at the same kinematical region at 70~GeV/c 
presented in the same figure. It supports our conclusion that asymmetry in 
quark scattering is not sensitive to the small energy difference in the 
initial state. 

We can conclude that the analyzing power in the inclusive $\pi^0$ 
production at high energies appears to illustrate the following features:

\begin{itemize}
\item {In the central region it is zero for reaction (2) 
[PROZA, E704, PHENIX] in the energy range $\sqrt{s}=10-200$~GeV and 
non zero for reaction (1) [PROZA only];}
\item {In the polarized particle fragmentation region for 
reactions (2) [PROZA, STAR] and (1) [PROZA] it is non zero and $A_N$ 
does not depend on the energy in the range $\sqrt{s}=10-200$~GeV for 
reaction (2).}
\end{itemize}

Current activity was partially supported by RFBR grant 06-02-16119.

\end{document}